\documentclass[epj]{webofc}
\usepackage[utf8]{inputenc}
\usepackage[varg]{txfonts}   
\usepackage{booktabs}
\usepackage{xcolor}
\usepackage{bbm}
\definecolor{darkred}{rgb}{0.4,0.0,0.0}
\definecolor{darkgreen}{rgb}{0.0,0.4,0.0}
\definecolor{darkblue}{rgb}{0.0,0.0,0.4}
\usepackage[bookmarks,linktocpage,colorlinks,
    linkcolor = darkred,
    urlcolor  = darkblue,
    citecolor = darkgreen]{hyperref}
\usepackage{subfigure}
\wocname{EPJ Web of Conferences}
\woctitle{Lattice2017}
\usepackage{float}
\restylefloat{table}
\begin{document}
%
\selectlanguage{english}
\title{%
Connected and disconnected contributions to nucleon axial form factors using $N_f=2$ twisted mass fermions at the physical point
}
\author{%
  \firstname{Constantia} \lastname{Alexandrou}\inst{1,2} \and
  \firstname{Martha} \lastname{Constantinou}\inst{3} \and
  \firstname{Kyriakos} \lastname{Hadjiyiannakou}\inst{2}\fnsep\thanks{Speaker, \email{k.hadjiyiannakou@cyi.ac.cy}} \and
  \firstname{Karl} \lastname{Jansen}\inst{4} \and
  \firstname{Christos} \lastname{Kallidonis}\inst{2} \and
  \firstname{Giannis} \lastname{Koutsou}\inst{2} \and
  \firstname{Alejandro} \lastname{Vaquero Avil\'es-Casco}\inst{5}
}
\institute{%
Department of Physics, University of Cyprus, P.O. Box 20537, 1678 Nicosia, Cyprus
\and
Computation-based Science and Technology Research Center, The Cyprus Institute, 20 Kavafi Str., Nicosia 2121, Cyprus
\and
Temple University,1925 N. 12th Street, Philadelphia, PA 19122-1801, USA
\and
NIC, DESY, Platanenallee 6, D-15738 Zeuthen, Germany
\and
Department of Physics and Astronomy, University of Utah, Salt Lake City, UT 84112, USA
}
\abstract{%
  We present results on the isovector and isoscalar nucleon axial form factors including disconnected contributions, using an ensemble of $N_f=2$ twisted mass clover-improved Wilson fermions simulated with approximately the physical value of the  pion mass. The light disconnected quark loops are computed using exact deflation, while the strange and the charm quark loops are evaluated using the truncated solver method. Techniques such as the summation and the two-state fits have been employed to access ground-state dominance.
}
\maketitle
\section{Introduction}\label{intro}

The form factors of the nucleon are important quantities that encapsulate information about its structure and properties. Contrary to the  electromagnetic form factors that are well determined experimentally, the axial form factors  are less known. The axial charge of the nucleon is an exception since it can
be  measured to high precision from $\beta$-decays. The momentum dependence of the axial form factors can be extracted from elastic scattering of neutrinos and protons \cite{Ahrens:1988rr}. The induced pseudoscalar form factor has been measured experimentally only for few values of momentum transfer~\cite{Choi:1993vt} from the cross section for exclusive $\pi^+$ electroproduction on the proton.

In this work, we evaluate the nucleon axial $G_A(Q^2)$ and induced pseudoscalar $G_p(Q^2)$ form factors using an ensemble of  $N_f=2$ twisted mass clover-improved Wilson ensemble with light quark mass tuned to approximately reproduce the physical value of the pion mass~\cite{Abdel-Rehim:2015pwa}. Both connected and disconnected contributions are evaluated allowing to compute the isovector, isoscalar as well as strange and charm form factors.

\section{Lattice Formulation}\label{sec-1}

\subsection{Axial form factors}\label{subsec-1.1}
The decomposition of the nucleon matrix element of the axial-vector current $A_\mu(x)$ in Euclidean time is given by 
\begin{equation}
\langle N(p',s') \vert A_\mu \vert N(p,s) \rangle = i \sqrt{ \frac{m_N^2}{E_N(\vec{p}') E_N(\vec{p})} } \bar{u}_N(p',s') \left( \gamma_\mu G_A(Q^2) - i \frac{Q_\mu}{2 m_N} G_p(Q^2) \right) \gamma_5 u_N(p,s),
\label{Eq:matrix_element}
\end{equation}
where  $p,s$ ($p',s'$) are the momentum and spin of the initial (final) nucleon state, $N$ is the nucleon state, $u_N$ the nucleon spinor, $m_N$ and $E_N(\vec{p})$ are the nucleon mass and energy with momentum $\vec{p}$ and   $Q^2=(p^\prime-p)^2$  
the momentum transfer square. 
We consider the isovector,  isoscalar as well as strange and charm combinations
\begin{equation}
  A^{\rm u-d}_\mu = \bar{\psi}(x) \gamma_\mu \gamma_5 \frac{\tau^3}{2} \psi(x), \,\, A^{\rm u+d}_\mu=\bar{\psi}(x) \gamma_\mu \gamma_5 \mathbbm{1} \psi(x), \,\, A^s_\mu = \bar{s}(x) \gamma_\mu \gamma_5 s(x)\,\,\,\textrm{and}\,\,\,A^c_\mu = \bar{c}(x) \gamma_\mu \gamma_5 c(x),
\end{equation}
where $\tau^3$ is the Pauli matrix acting in flavor space.

\subsection{Lattice extraction}\label{subsec-1.2}
Computation of two- and three-point correlation functions is needed to extract nucleon matrix elements. The three-point functions receive contributions from the so-called connected  and disconnected diagrams. For the isovector combination disconnected contributions cancel out in the isospin limit. For the connected contributions we employ the standard fixed-sink method where sequential inversions through the sink are performed. Deflation of the low modes is employed to accelerate the inversion of the Dirac operator. For the disconnected quark loops we combined the one-end trick \cite{McNeile:2006bz} with the truncated solver method (TSM) \cite{Bali:2009hu} to reduce the computational cost. Two-point functions are computed for several source positions per configuration to increase the statistical accuracy.

We construct the following ratio of the appropriate three-point function $G_\mu(\Gamma_\nu,\vec{p}\,',\vec{p};t_s,t_{\rm ins})$ to two-functions $C(\Gamma_0,\vec{p},t)$
\begin{equation}
R_\mu(\Gamma_\nu,\vec{p}\,',\vec{p};t_s,t_{\rm ins}) = \frac{G_\mu(\Gamma_\nu,\vec{p}\,',\vec{p};t_s,t_{\rm ins})}{C(\Gamma_0,\vec{p}\,';t_s)} \sqrt{\frac{C(\Gamma_0,\vec{p};t_s-t_{\rm ins}) C(\Gamma_0,\vec{p}\,';t_{\rm ins}) C(\Gamma_0,\vec{p}\,';t_s)}{C(\Gamma_0,\vec{p}\,';t_s-t_{\rm ins}) C(\Gamma_0,\vec{p};t_{\rm ins}) C(\Gamma_0,\vec{p};t_s)}},
\label{Eq:ratio}
\end{equation}
which, in the large time limit $t_s-t_{\rm ins} \gg 1$ and $t_{\rm ins} \gg 1$
yields the desired nucleon matrix element. The insertion time $t_{\rm ins}$ as well as the sink time $t_s$ are taken relative to the source. To determine, if indeed, these  time separations are large enough  we employ three methods: i) \textit{plateau method}, which assumes that the ratio of Eq.~(\ref{Eq:ratio}) is dominated by  the ground state and perform a constant fit as a function of $t_{\rm ins}$  to extract the matrix element, ii) \textit{summation method}, in which one sums over $t_{\rm ins}$  and extracts the matrix element from the slope of a linear fit, and iii) \textit{two-state fit method} which includes besides the ground state the first excited state in the fit to extract the matrix element. We require that these three methods give consistent results for the matrix element.

 We perform a non-perturbative calculation of both the  renormalization functions for the isovector and isoscalar currents needed for the extraction of physical matrix elements from lattice results using the Rome-Southampton method \cite{Martinelli:1994ty}. Lattice artifacts are subtracted perturbatively to ${\cal O}(a^2)$~\cite{Alexandrou:2015sea} to yield a better determination of the limit $(ap)^2 \rightarrow 0$. We take the chiral limit  to extract the renormalization functions. We find for the non-singlet case $Z_A^{ns}=0.7910(4)(5)$ and for the singlet $Z_A^s =0.7968(25)(91)$, which are compatible within errors.

\subsection{Lattice Setup and Statistics}\label{subsec-1.4}
We analyze an ensemble of $N_f=2$ twisted mass clover improved Wilson fermions with pion mass $m_\pi=0.1304(4)$~GeV on a lattice of size $48^3\times 96$ and a lattice spacing $a$=0.0938(3)~fm determined from nucleon mass \cite{Alexandrou:2017xwd}. In Tab.~\ref{Tab:statistics} we give the statistics used for the computation of connected and disconnected contributions.  For the connected, three values of $t_s$ analyzed in the frame where $\vec{p}^\prime=\vec{0}$, while for the disconnected all separations are available without additional cost in both the rest frame of the final nucleon as well as for  $\vec{p}^\prime = \frac{2\pi\hat{\vec{n}}}{L}$. We note the much larger number of configurations and source positions analyzed for the evaluation of the disconnected contributions.

\begin{table}[H]
  \begin{center}
\renewcommand{\arraystretch}{1.2}
\renewcommand{\tabcolsep}{3.5pt}
\begin{tabular}{ccc||ccccc}
\multicolumn{3}{c||}{Connected} & \multicolumn{5}{c}{Disconnected} \\
\hline\hline
$t_s/a$ & $N_{\rm conf}$ & $N_{\rm src}$ & Flavor & $N_{\rm conf}$ & $N_r^{\rm HP}$ & $N_r^{\rm LP}$ & $N_{\rm src}$ \\
\hline
10 & 579 & 16 & light   & 2120 & 2250 &   -  & 100 \\
12 & 579 & 16 & strange & 2057 & 63   & 1024 & 100 \\
14 & 579 & 16 & charm   & 2034 & 5    & 1250 & 100 \\
\hline
\end{tabular}
\vspace{-0.5cm}\caption{Statistics used in this study. $N_{\rm conf}$ is the number of gauge configurations and $N_{\rm src}$ is the number of source positions per configuration. For the disconnected contributions, $N_r^{\rm HP}$ is the number of high-precision stochastic vectors produced, and $N_r^{\rm LP}$ is the number of low-precision vectors used when the TSM has been used.} 
\label{Tab:statistics}
\end{center}
\end{table}

\section{Results}\label{sec-2}

\subsection{Axial charge}\label{subsec-2.1}
\begin{figure}[!ht]\vspace*{-1cm}
  \begin{center}
    \begin{minipage}[t]{0.48\linewidth}
      \vspace*{-6.5cm}
    \includegraphics[width=\textwidth]{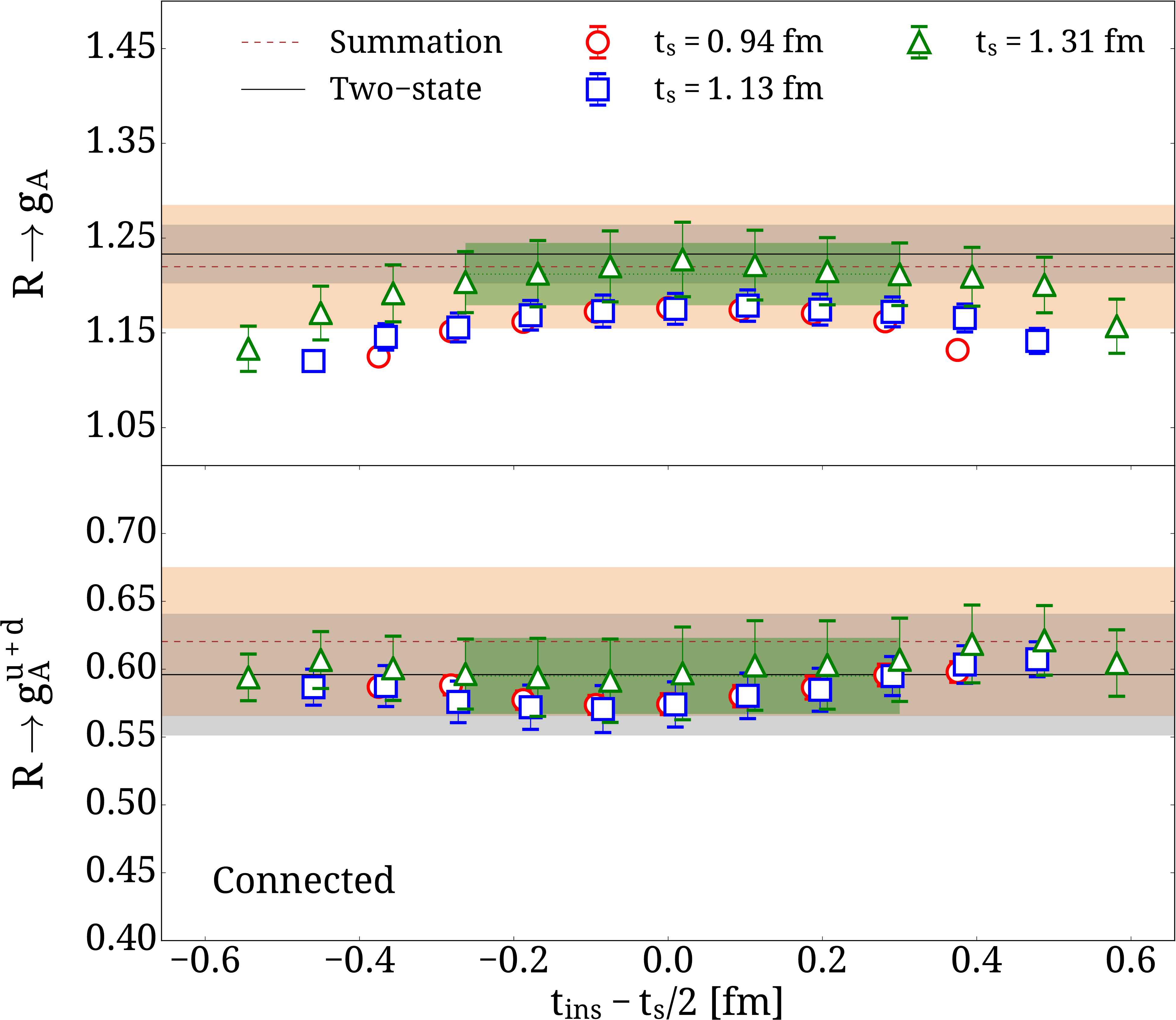}
  \end{minipage}
  \hfill
  \begin{minipage}[t]{0.48\linewidth}
    \includegraphics[width=\textwidth]{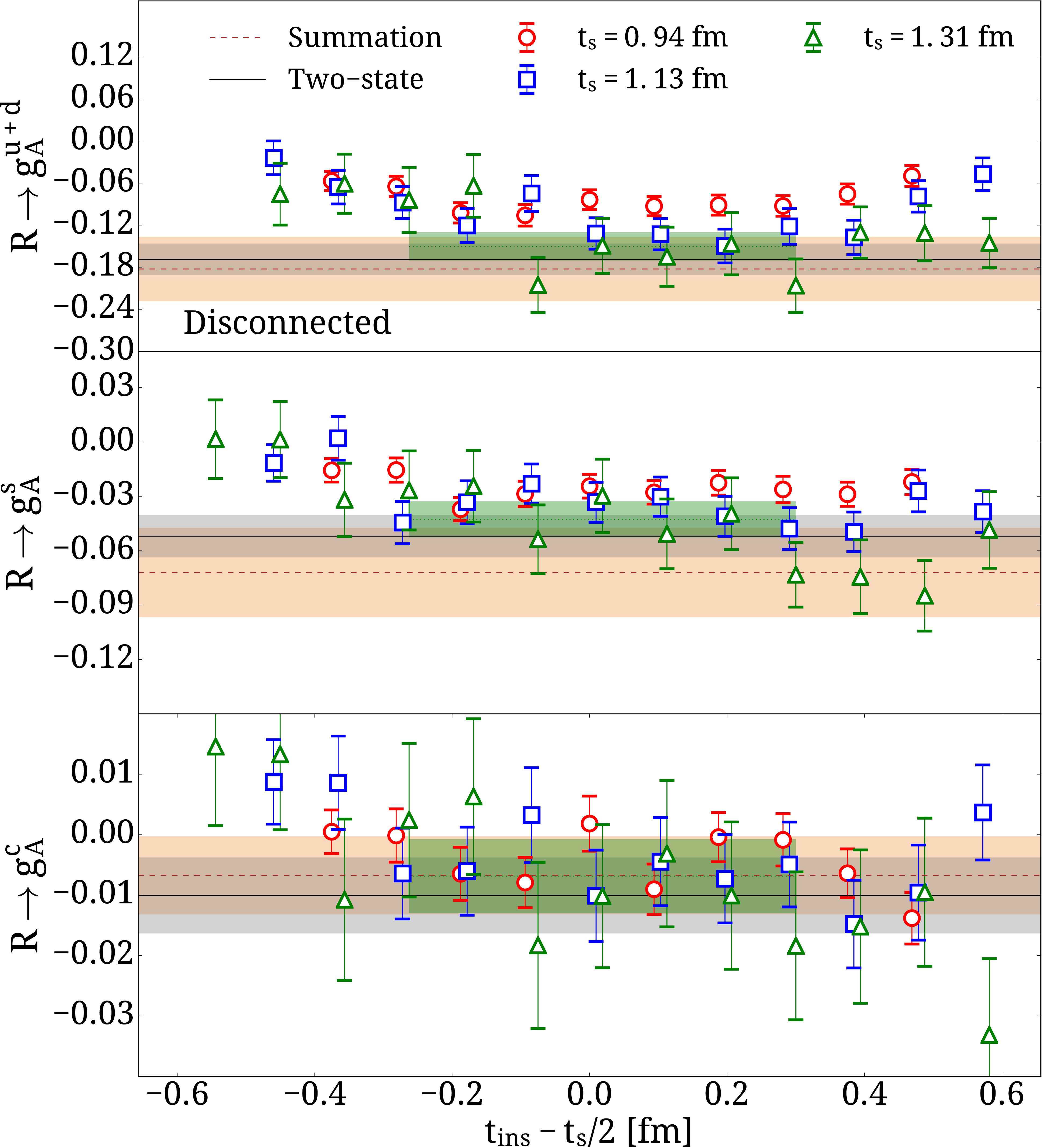}
  \end{minipage}
  \vspace{-0.3cm}\caption{Left: Ratio from where we extract $g_A$ and the connected contributions to  $g_A^{u+d}$. Results are presented for three values of $t_s$, namely $t_s=0.94,\; 1.13$ and 1.31~fm shown with open red circles, open blue squares and open green triangles, respectively. The constant fit using the plateau method is shown with the dotted line spanning the selected fit range of $t_{\rm ins}$ and its corresponding error band. Results extracted from the summation method are shown with the brown dashed line and corresponding error band, while results using two-state fits are shown with the solid black line spanning the entire horizontal axis. Right: Ratio from where we extract the disconnected contributions to $g_A^{u+d}$, $g_A^{s}$ and $g_A^c$. The convention is as the left panel.} 
  \label{Fig:ratios_gA}
\hfill
\end{center}  
\end{figure}

At zero momentum transfer the matrix element of the axial-vector current  yields the nucleon axial charge $g_A \equiv g_A^{u-d}$ (isovector) and $g_A^{u+d}$ (isoscalar). In Fig.~\ref{Fig:ratios_gA} we present our results for both quantities showing separately the connected and disconnected contributions in the case of $g_A^{u+d}$. The ratio for the connected contributions is computed at three values of $t_s$. We check that the values extracted using the plateau, summation and two-state methods are consistent as shown in Fig.~\ref{Fig:ratios_gA}.
We include a systematic error due to the excited states taken as the difference between the plateau value that demonstrates convergence with $t_s$ and the one extracted from the two-state fit. We quote these  values  in Tab.~\ref{Tab:Val_gA}.
\begin{table}[!ht]
  \begin{center}
    \small\addtolength{\tabcolsep}{-3pt}
    \begin{tabular}{| c | c | c | c | c | c | c | c }
      \hline
      $g_A$ & $g_A^{u+d}$ (Conn.) & $g_A^{u+d}$ (Disc.) & $g_A^{u+d}$ & $g_A^s$ & $g_A^c$ \\
      \hline\hline
      1.212(33)(22) & 0.595(28)(1) & -0.150(20)(19) & 0.445(34)(19) & -0.0427(100)(93) & -0.00338(188)(667) \\
      \hline
    \end{tabular}
  \end{center}
  \vspace*{-0.5cm} \caption{Our values for the nucleon axial charges. The first error is statistical and the second is a systematic due to the excited states contamination.}
  \label{Tab:Val_gA}
\end{table}

\subsection{Isovector Axial and induced pseudoscalar form factors}\label{subsec-2.2}
\begin{figure}[!ht]\vspace*{-1cm}
\begin{center}
  \begin{minipage}[t]{0.47\linewidth}
    \includegraphics[width=\textwidth]{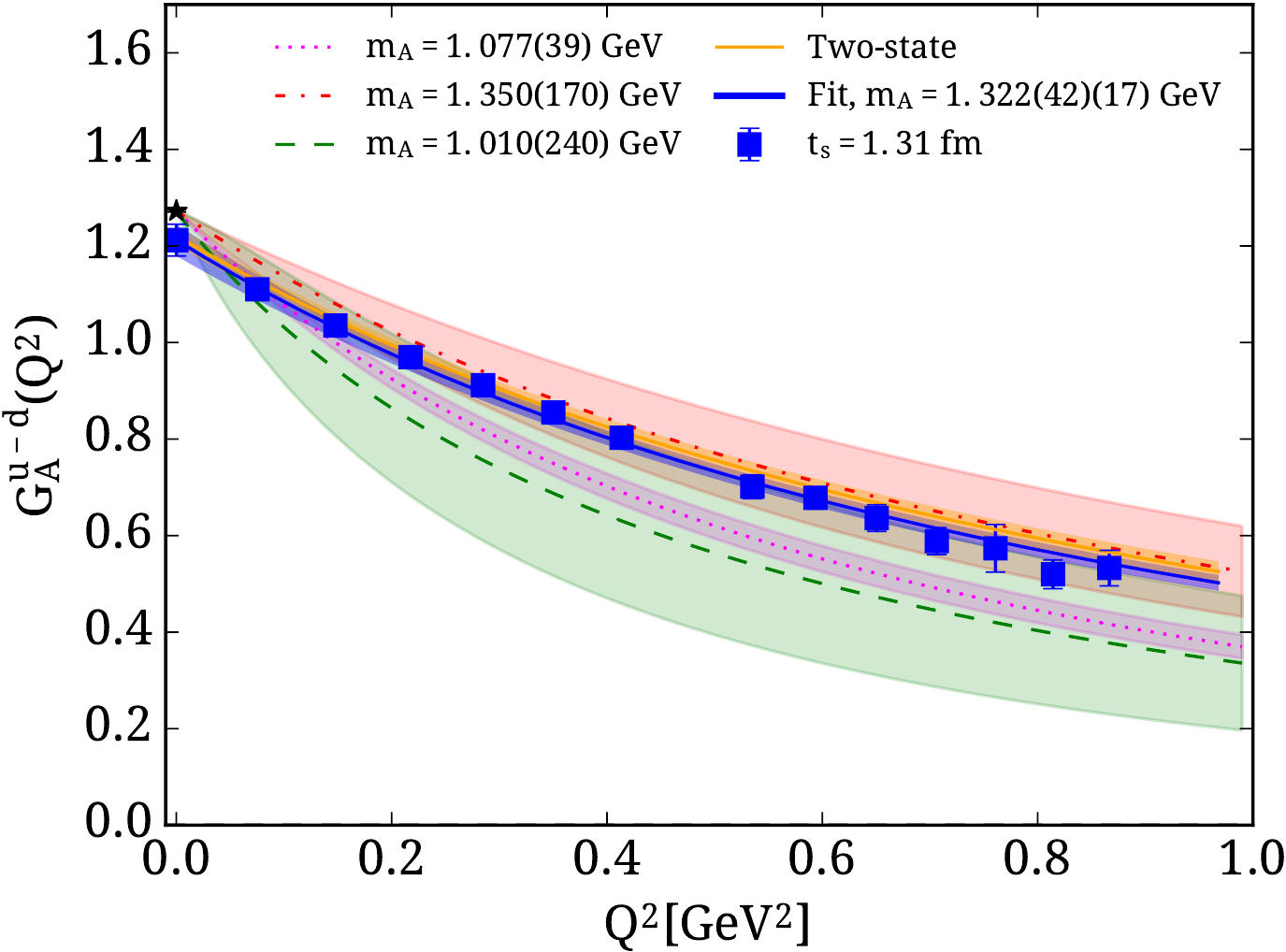}
  \end{minipage}
  \hfill
  \begin{minipage}[t]{0.47\linewidth}
    \includegraphics[width=\textwidth]{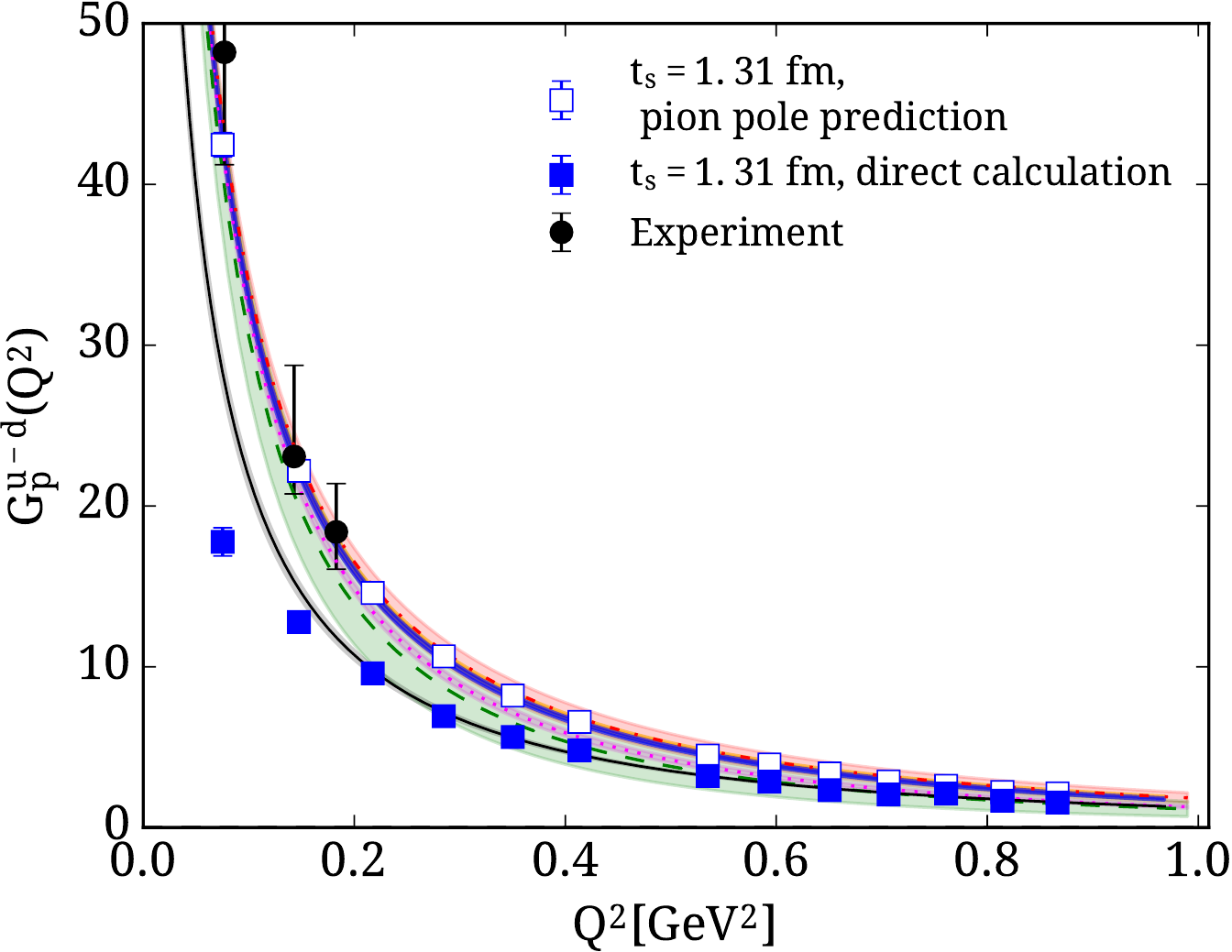}
  \end{minipage}
\hfill
\vspace*{-0.3cm}\caption{
  Results for $G_A^{u-d}(Q^2)$ and $G_p^{u-d}(Q^2)$ 
  using values extracted from the plateau method at $t_s=1.31$~fm (filled blue squares). In the left panel,  the solid blue (orange) line shows the fit to our results extracted from the plateau (two-state fit) using dipole form. Black asterisk shows the experimental value of $g_A$. The purple, red and green bands are experimental results for $G_A^{u-d}(Q^2)$ taken from Refs.~\cite{Liesenfeld:1999mv}, \cite{AguilarArevalo:2010zc} and~\cite{Meyer:2016oeg} respectively.
For the induced pseudoscalar, the open blue squares show the pion-pole prediction results to $G_p^{u-d}(Q^2)$ from our lattice results of $G_A^{u-d}(Q^2)$ together with the corresponding fits, blue (orange) band is a fit extracted from  the plateau (two-state) fit. The filled blue squares show  $G_p^{u-d}(Q^2)$ extracted directly from the nucleon matrix element with a pole fit (solid black line)  after omitting the two lowest $Q^2$ values. The filled black circles are experimental results of $G_p^{u-d}(Q^2)$ from Ref.~\cite{Choi:1993vt}. The purple, red and green bands are constructed using the experimental results for $G_A^{u-d}(Q^2)$ and pion pole to infer $G_p^{u-d}(Q^2)$. }
\label{Fig:GA_Gp_bands}
\end{center}  
\end{figure}

In Fig.~\ref{Fig:GA_Gp_bands} we show results for the isovector  $G_A$ and $G_p$. For the axial form factor we fit our results to a dipole form: $G_A(Q^2) = \frac{g_A}{(1+Q^2/m_A^2)^2}$ where $g_A$ is fixed by the value of the form factor at zero momentum transfer and the axial mass $m_A$ is allowed to vary. We find an axial mass $m_A=1.322(42)(17)$~GeV, which is consistent with the recent experimental value \cite{AguilarArevalo:2010zc} from MiniBooNE experiment but larger than the one extracted from previous experiments. For the induced pseudoscalar form factor we fit to a pole form $G_p(Q^2)=G_A(Q^2) \frac{C}{Q^2+m_p^2}$ where $C$ and $m_p$ are fit parameters. As shown in Fig.~\ref{Fig:GA_Gp_bands} our lattice results display a  milder $Q^2$-dependence compared to the one expected from the pion pole dominance. This discrepancy at low $Q^2$ values might be due to volume  effects that suppress pion cloud formation and need to be investigated in future studies using bigger volumes and better interpolating fields.

\subsection{Disconnected contributions to the axial form factors}\label{subsec-2.3}

\begin{figure}[!ht]
\begin{center}
  \begin{minipage}[t]{0.47\linewidth}
    \includegraphics[width=\textwidth]{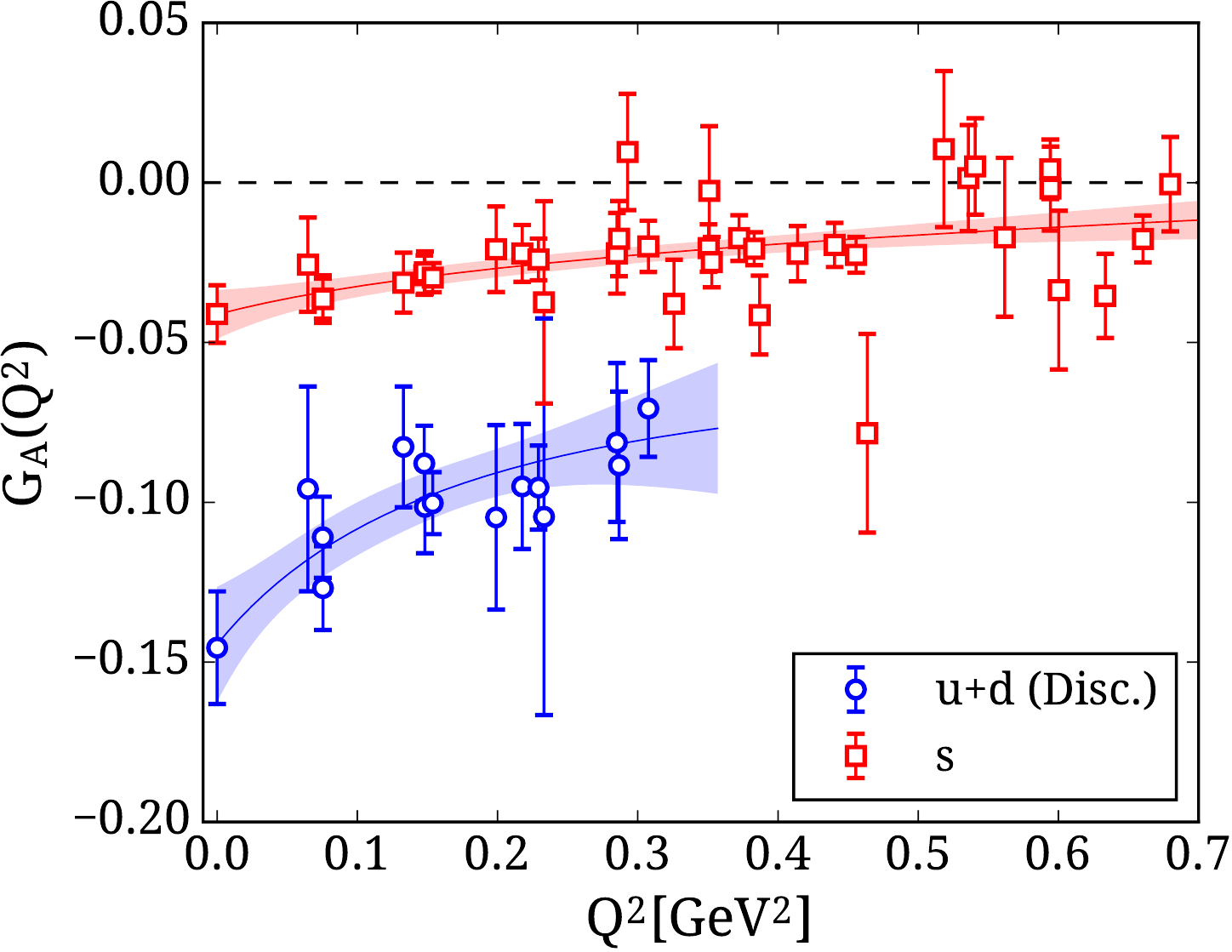}
  \end{minipage}
  \hfill
  \begin{minipage}[t]{0.47\linewidth}
    \includegraphics[width=\textwidth]{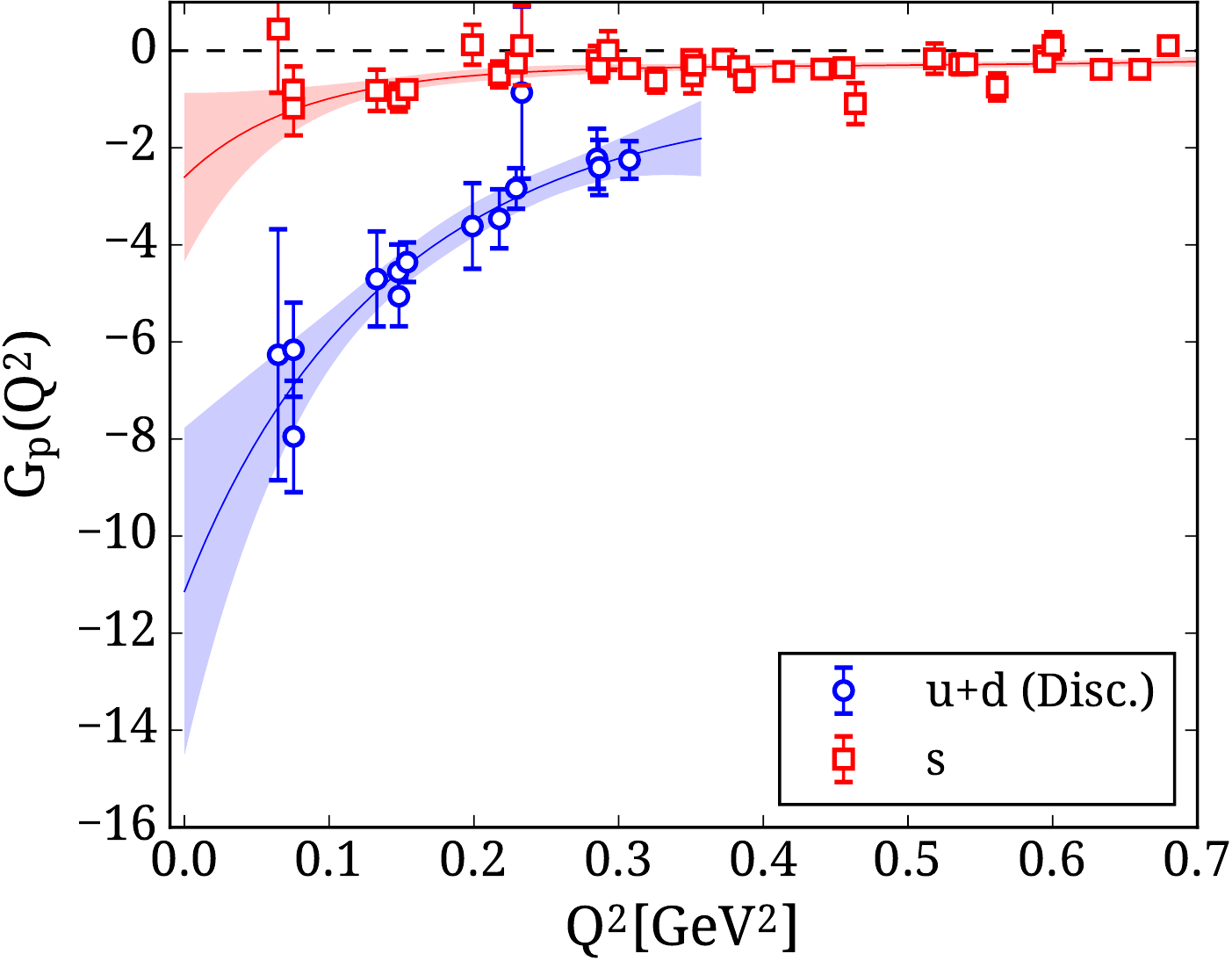}
  \end{minipage}
\hfill
\vspace*{-0.3cm}\caption{Results for the disconnected contributions: Left $G_A^{u+d}(Q^2)$ (red)  and $G_A^s(Q^2)$ (blue. Right: $G_p(Q^2)^{u+d}$ (red) and $G_p^s(Q^2)$ (blue).
  The lattice  results are extracted from the plateau method at $t_s=1.31$~fm. The bands show the fit using the z-expansion.}
\label{Fig:GA_Gp_disc}
\end{center}  
\end{figure}

In Fig.~\ref{Fig:GA_Gp_disc} we show the disconnected contributions to  the isoscalar $G_A^{u+d}$ and $G_p^{u+d}$ as well as the strange  $G_A^s$ and $G_p^s$
form factors. We perform a model independent fit to these data using the z-expansion \cite{Hill:2010yb} that yields a good description of the $Q^2$-dependence.

\begin{figure}[!ht]
  \begin{center}
    \begin{minipage}[t]{0.47\linewidth}
      \mbox{}\\[-\baselineskip]
      \includegraphics[width=\textwidth]{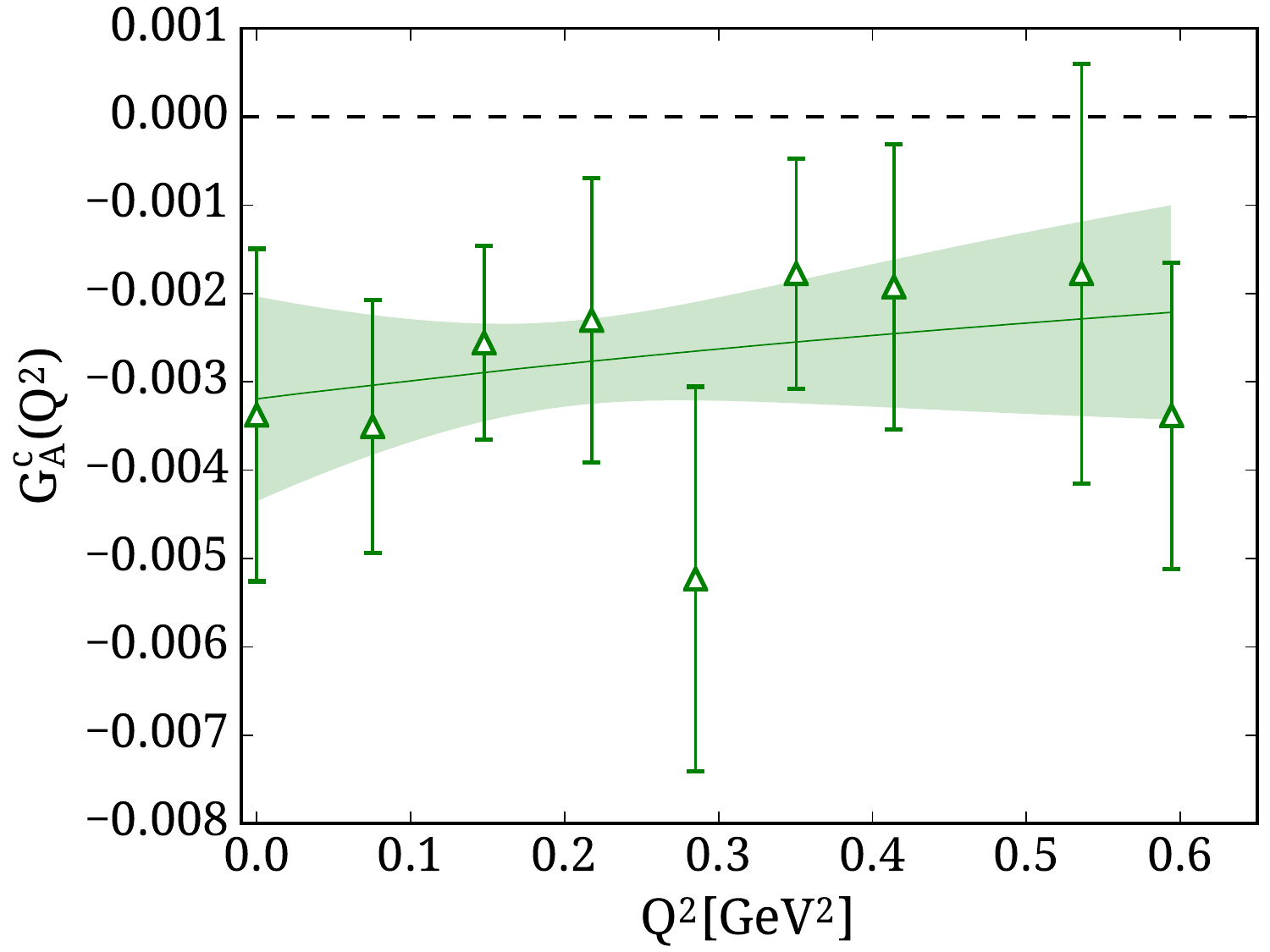}
    \end{minipage}
    \hfill
    \begin{minipage}[t]{0.47\linewidth}
      \mbox{}\\[-\baselineskip]
      \caption{Results for $G_A^c(Q^2)$ extracted using the plateau method at $t_s=0.94$~fm. The fit is performed using a dipole form. }
      \label{Fig:GA_c}
    \end{minipage}
  \end{center}
\end{figure}

In Fig.~\ref{Fig:GA_c} we show our results for $G_A^c(Q^2)$ extracted from the plateau method for a source-sink time separation $t_s=0.94$~fm. Due to the fact that the statistical uncertainty for this quantity is large it is not possible to study larger separations. $G_A^c(Q^2)$ is clearly negative and non-zero, with values that are an order of magnitude smaller as compared to $G_A^s(Q^2)$. The z-expansion requires high accuracy 
and thus we fit the $Q^2$-dependence using a dipole form. $G_p^c(Q^2)$ is very noisy to display and it is omitted.
\vspace{0.5cm}

\subsection{Isoscalar Axial and induced pseudoscalar form factors}\label{subsec-2.4}
\begin{figure}[!ht]
\begin{center}
  \begin{minipage}[t]{0.47\linewidth}
    \includegraphics[width=\textwidth]{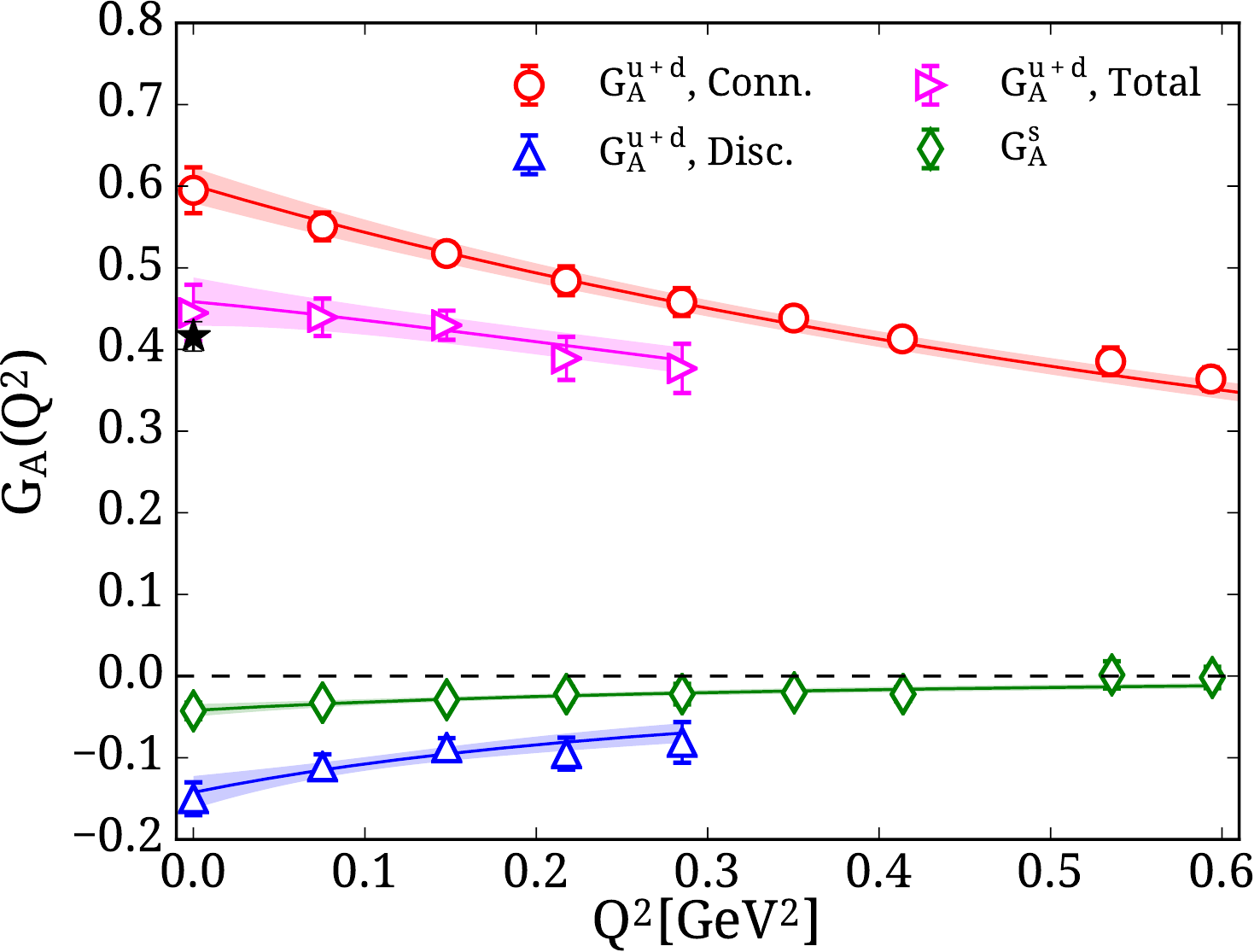}
  \end{minipage}
  \hfill
  \begin{minipage}[t]{0.47\linewidth}
    \includegraphics[width=\textwidth]{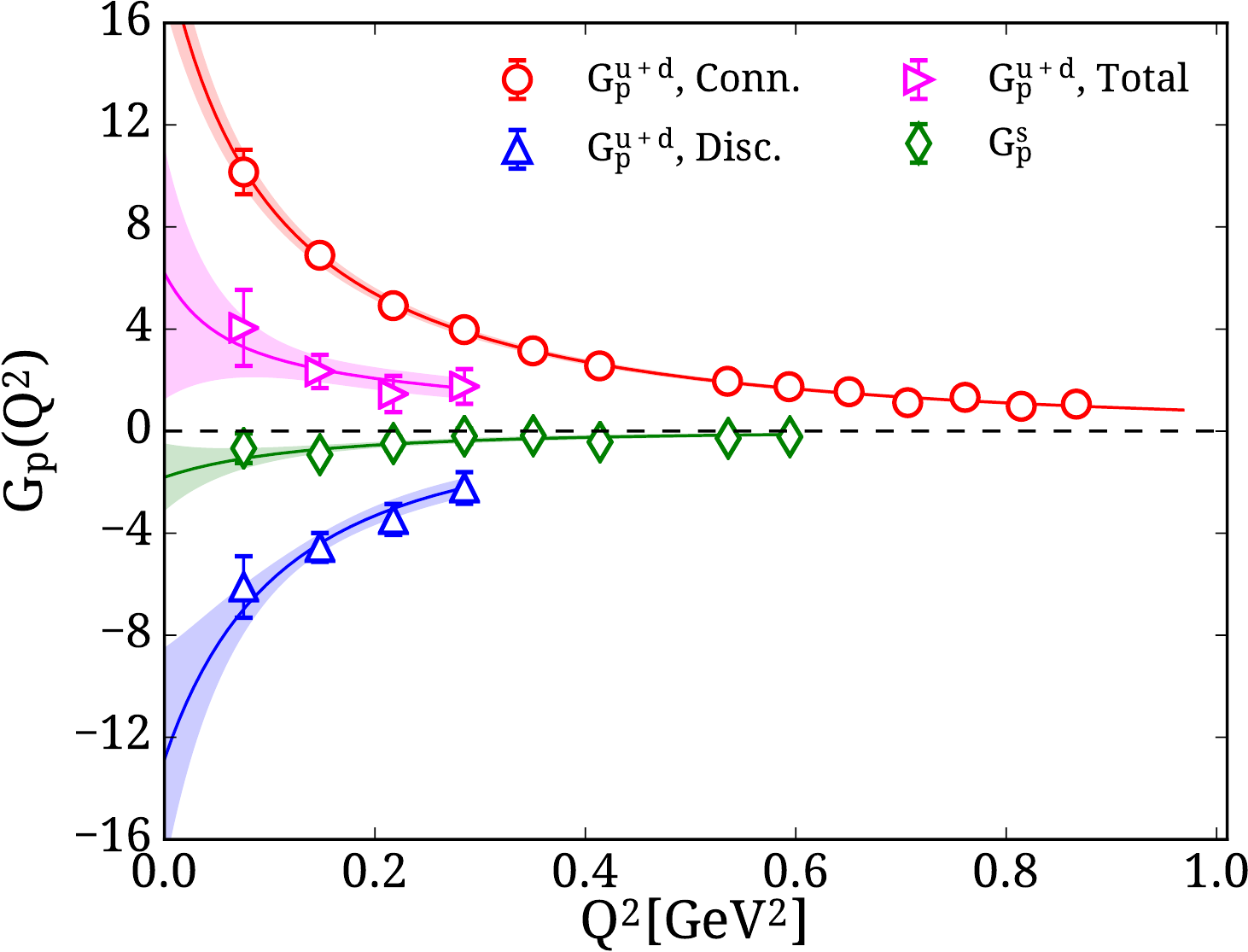}
  \end{minipage}
\hfill
\vspace*{-0.3cm}\caption{Results for the isoscalar $G_A^{u+d}(Q^2)$ and $G_A^s(Q^2)$ (left) and $G_p^{u+d}(Q^2)$ and $G_p^s(Q^2)$ (right). Connected contributions to the isoscalar form factors are shown by the red open circles, disconnected by the  blue upper open triangles, and the total by the right magenta open triangles. The strange form factors are shown by the green rhombus.  The black star 
shows the experimental value of $g_A^{u+d}$. }
\label{Fig:GA_Gp_isos}
\end{center}  
\end{figure}

Having both connected and disconnected contributions allows us to compute the total contribution to the isoscalar form factor. In  Fig.~\ref{Fig:GA_Gp_isos} we show results for the isoscalar $G_A^{u+d}$. As can be seen, the disconnected contribution comes with a different sign compared to the connected one, it is clearly non-zero and changes the $Q^2$-dependence of the form factor. Only after we include the disconnected contribution 
we have agreement with the experimental value of $g_A^{u+d}$. The isoscalar mass $m_A^{u+d}=1.736(244)(374)$  is higher than the one extracted from the isovector case as expected from the smoother $Q^2$-dependence observed of the isoscalar  form factor.

Our results on the isoscalar $G_p^{u+d}$  are presented in Fig.~\ref{Fig:GA_Gp_isos}. What is remarkable is the large disconnected contribution
to the induced pseudoscalar form factor, which is of the same order
as the connected but with the opposite sign. Fitting separately the two contributions we find consistent pole masses, namely $m_p^{\rm u+d,\; conn}=0.324(22)(12)$~GeV and $m_p^{\rm u+d,\; disc}=0.331(81)(36)$ showing that the two contributions are dominated by the same pole mass canceling the pion mass dependence in the isoscalar form factor.

\subsection{Comparison with other recent lattice QCD results}
\begin{figure}
\begin{center}
  \begin{minipage}[t]{0.47\linewidth}
    \includegraphics[width=\textwidth]{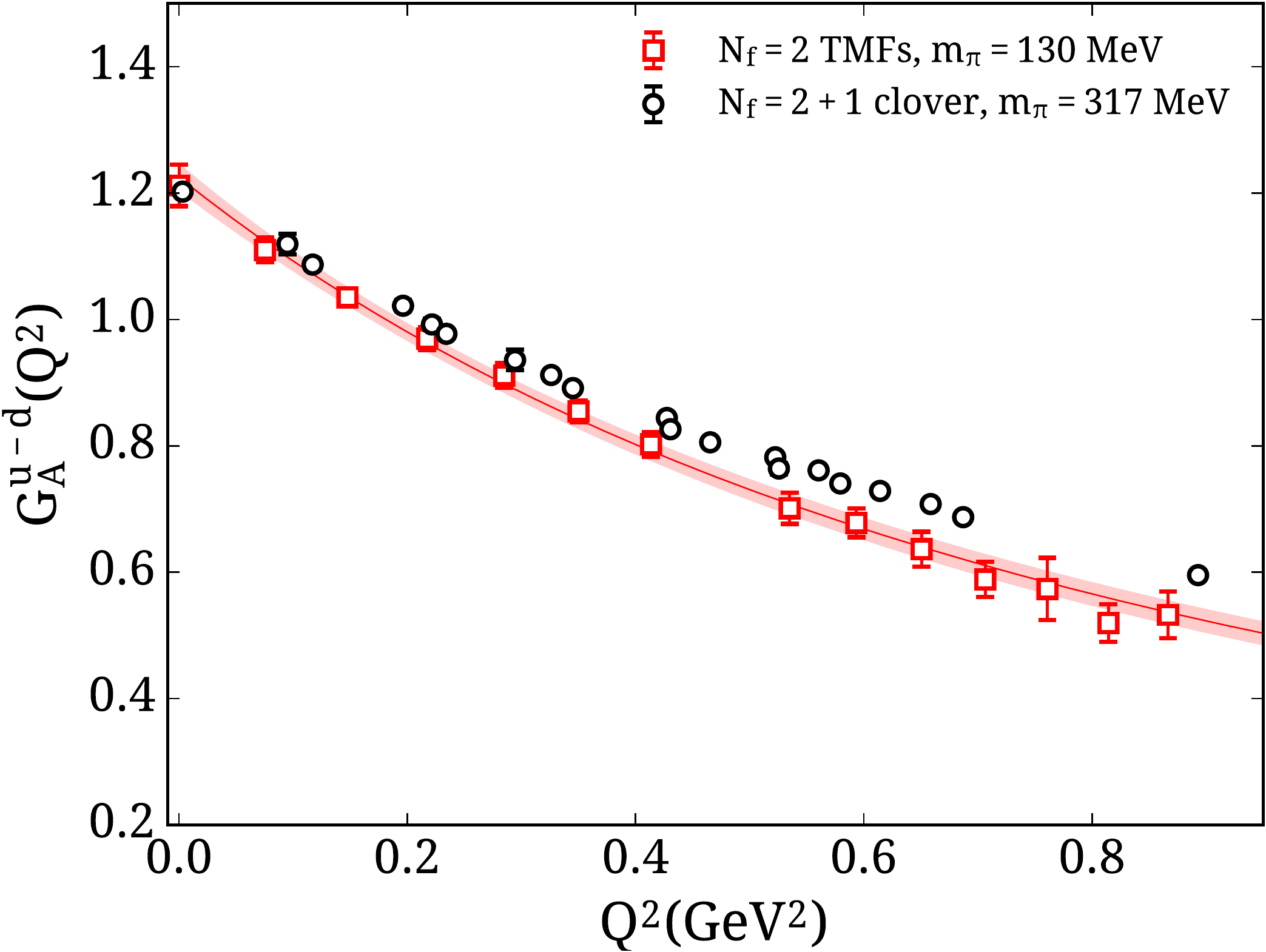}
  \end{minipage}
  \hfill
  \begin{minipage}[t]{0.47\linewidth}
    \includegraphics[width=\textwidth]{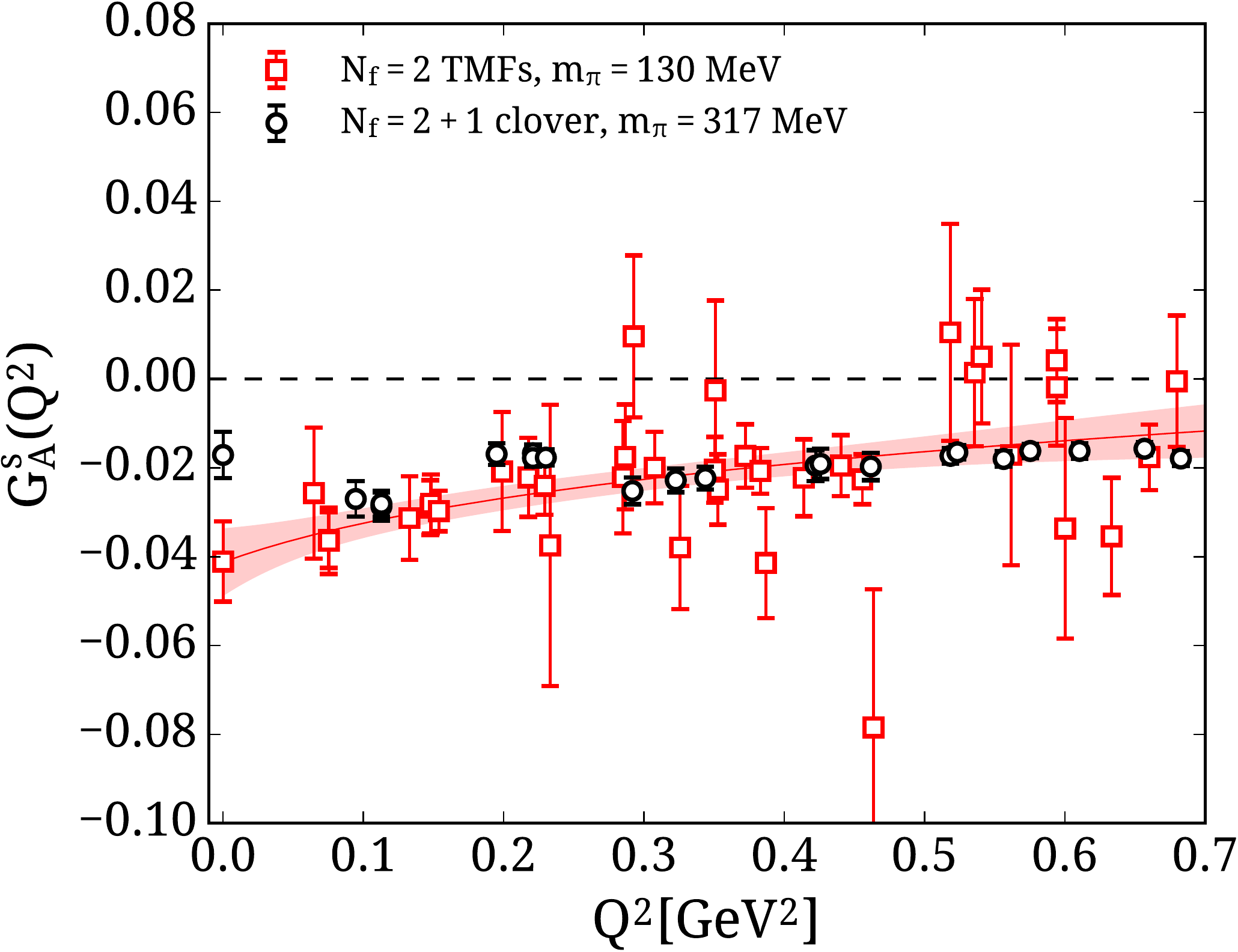}
  \end{minipage}
\hfill
\vspace*{-0.3cm}\caption{Comparison of our results on $G_A^{u-d}(Q^2)$ (left) and $G_A^s(Q^2)$ (right) using twisted mass fermions (TMFs) with approximately a physical value of the light quark mass~\cite{Alexandrou:2017hac} (open red squares) with results using $N_f=2+1$ clover fermions ~\cite{Green:2017keo} with light quark mass yielding pion mass of 317 MeV (open black circles). The red band shows the fit to our results with its jackknife error.}
\label{Fig:comparison}
\end{center}  
\end{figure}

In Fig.~\ref{Fig:comparison} we show a comparison of our results at the physical point with results from another study at a higher pion mass that used $N_f=2+1$ clover-improved Wilson fermions \cite{Green:2017keo} with pion mass of 317~MeV. As can be seen, the values of   $G_A^{u-d}(Q^2)$ 
are consistently larger as compared to ours in particular at larger values of $Q^2$ leading to a milder slope than ours. Assuming that lattice artifacts are small in both calculations this indicates  a non-zero pion mass dependence.
There is an overall agreement between results on $G_A^s(Q^2)$  Fig.~\ref{Fig:comparison} with the exception of the value at $Q^2=0$. We  note that the results
from Ref.~\cite{Green:2017keo} used a smaller source-sink time separation. This together with the fact that the pion mass is larger than physical is reflected in having smaller statistical errors even though  our statistics are twice as large. 

\section{Conclusions}
The isovector, isoscalar, strange and charm axial and induced pseudoscalar form factors of the nucleon are presented using an  ensemble of two degenerate dynamical quarks with mass tuned to approximately reproduce the physical value of the pion mass (physical point)~\cite{Alexandrou:2017hac}. Disconnected contributions are evaluated for the first time at the physical point using improved methods providing accurate results. They are shown to be non-zero and of opposite sign to the connected contribution. For the case of the induced pseudoscalar form factor the disconnected are of the same magnitude as the connected canceling the pion pole in the isoscalar $G_p^{u+d}(Q^2)$.
Only after adding the disconnected contribution to the isoscalar axial charge
we have agreement with the experimental value. Additionally, $G_A^s(Q^2)$ and $G_A^c(Q^2)$ are found to be non-zero and  negative, as is $G_p^s(Q^2)$, while $G_p^c(Q^2)$ is very noisy.

The value of the nucleon axial charge is lower by one standard deviation as compared to the experimental value and the slope of $G_A^{u-d}(Q^2)$
is milder but  in agreement with the experimental result from Ref.~\cite{AguilarArevalo:2010zc}. However, one needs to study the $Q^2$-dependence further since
lattice QCD results tend to underestimate the slope.  $G_p^{u-d}(Q^2)$ also displays a milder $Q^2$-dependence than expected from pion pole dominance. We are currently investigating volume effects on these quantities using an ensemble with the same parameters as the one analyzed here but with a lattice size of $64^3\times 128$.

\emph{Acknowledgments}: We acknowledge funding from the European Union’s Horizon 2020 research and innovation programme under the Marie Sklodowska-Curie grant agreement No 642069. This work was partly supported by a grant from the Swiss National Supercomputing Centre (CSCS) under project IDs s540 and s625 on the Piz Daint system, by a Gauss allocation on SuperMUC with ID 44060 and in addition used computational resources from the John von Neumann-Institute for Computing on the Jureca and the BlueGene/Q Juqueen systems at the research center in Julich. We also acknowledge PRACE for awarding us access to the Tier-0 computing resources Curie, Fermi and SuperMUC based in CEA, France, Cineca, Italy and LRZ, Germany, respectively. K. H. and Ch. K. acknowledge support from the Cyprus Research Promotion Foundation under contract T$\Pi$E/$\Pi\Lambda$HPO/0311(BIE)/09. We also acknowledge financial support from the PRACE-4IP project with grant number 653838.

\clearpage
\bibliography{lattice2017}

\begin{thebibliography}{14}

\bibitem{Ahrens:1988rr}
L.A. Ahrens et~al., Phys. Lett. \textbf{B202}, 284 (1988)

\bibitem{Choi:1993vt}
S.~Choi et~al., Phys. Rev. Lett. \textbf{71}, 3927 (1993)

\bibitem{Abdel-Rehim:2015pwa}
A.~Abdel-Rehim et~al. (ETM), Phys. Rev. \textbf{D95}, 094515 (2017),
  \texttt{1507.05068}

\bibitem{McNeile:2006bz}
C.~McNeile, C.~Michael (UKQCD), Phys. Rev. \textbf{D73}, 074506 (2006),
  \texttt{hep-lat/0603007}

\bibitem{Bali:2009hu}
G.~Bali, S.~Collins, A.~Schafer, Comput. Phys. Commun. \textbf{181}, 1570
  (2010), \texttt{0910.3970}

\bibitem{Martinelli:1994ty}
G.~Martinelli, C.~Pittori, C.T. Sachrajda, M.~Testa, A.~Vladikas, Nucl. Phys.
  \textbf{B445}, 81 (1995), \texttt{hep-lat/9411010}

\bibitem{Alexandrou:2015sea}
C.~Alexandrou, M.~Constantinou, H.~Panagopoulos (ETM), Phys. Rev. \textbf{D95},
  034505 (2017), \texttt{1509.00213}

\bibitem{Alexandrou:2017xwd}
C.~Alexandrou, C.~Kallidonis, Phys. Rev. \textbf{D96}, 034511 (2017),
  \texttt{1704.02647}

\bibitem{AguilarArevalo:2010zc}
A.A. Aguilar-Arevalo et~al. (MiniBooNE), Phys. Rev. \textbf{D81}, 092005
  (2010), \texttt{1002.2680}

\bibitem{Liesenfeld:1999mv}
A.~Liesenfeld et~al. (A1), Phys. Lett. \textbf{B468}, 20 (1999),
  \texttt{nucl-ex/9911003}

\bibitem{Meyer:2016oeg}
A.S. Meyer, M.~Betancourt, R.~Gran, R.J. Hill, Phys. Rev. \textbf{D93}, 113015
  (2016), \texttt{1603.03048}

\bibitem{Hill:2010yb}
R.J. Hill, G.~Paz, Phys. Rev. \textbf{D82}, 113005 (2010), \texttt{1008.4619}

\bibitem{Green:2017keo}
J.~Green, N.~Hasan, S.~Meinel, M.~Engelhardt, S.~Krieg, J.~Laeuchli, J.~Negele,
  K.~Orginos, A.~Pochinsky, S.~Syritsyn, Phys. Rev. \textbf{D95}, 114502
  (2017), \texttt{1703.06703}

\bibitem{Alexandrou:2017hac}
C.~Alexandrou, M.~Constantinou, K.~Hadjiyiannakou, K.~Jansen, C.~Kallidonis,
  G.~Koutsou, A.~Vaquero Aviles-Casco, Phys. Rev. \textbf{D96}, 054507 (2017),
  \texttt{1705.03399}

\end{thebibliography}

\end{document}